\documentclass[nofootinbib]{revtex4}
\usepackage{graphicx}
\usepackage{latexsym}
\def\be{\begin{equation}}
\def\ee{\end{equation}}
\def\bea{\begin{eqnarray}}
\def\eea{\end{eqnarray}}
\def\bi#1{\hbox{\boldmath{$#1$}}}

\begin{document}

\title{New Constraints on Anisotropic Rotation of CMB Polarization}

\author{Mingzhe Li$^{1}$}
\email{limz@ustc.edu.cn}
\author{Bo Yu$^{2,3}$ }
\email{yubo@pmo.ac.cn}
\affiliation{${}^1$
Interdisciplinary Center for Theoretical Study, University of Science and Technology of China, Hefei, Anhui 230026, China}
\affiliation{${}^2$
Purple Mountain Observatory, Chinese Academy of Sciences, China}
\affiliation{${}^3$ Joint Center for Particle, Nuclear
Physics and Cosmology, Nanjing University - Purple Mountain
Observatory, Nanjing 210093, China}

%\date{\today.}

\begin{abstract}

The coupling of a scalar field to electromagnetic field via the Chern-Simons term will rotate the polarization directions of the
cosmic microwave background radiation. The rotation angle which relies on the distribution of the scalar field on the CMB sky is
direction dependent. Such anisotropies will give rise to new distortions to the power spectra of CMB polarization and it can be used to
probe the detailed physics of the scalar field. In this paper we use the updated observational data to constrain the anisotropic rotation
angle in a model independent way. We find that the dominant effect of the anisotropic rotation on CMB comes from its variance and it is
constrained tightly by the current data.

\end{abstract}

\maketitle
\hfill USTC-ICTS-13-03
%\hskip 1.6cm PACS number(s): 98.80.Es, 98.80.Cq, 98.70.Vc \vskip 0.4cm

\section{Introduction}

As is well known the coupling of the electromagnetic field to a given
external field via the Chern-Simons term $p_{\mu}A_{\nu}\widetilde{F}^{\mu\nu}$ in the Lagrangian density will give rise to
the rotation of linear polarization of the photons as they propagate in the background spacetime in which $p_{\mu}$
does not vanish \cite{cfj}.
Here $A_{\mu}$ is the electromagnetic vector field, the antisymmetric tensor of the electromagnetic field is $F_{\mu\nu}=
\partial_{\mu}A_{\nu}-\partial_{\nu}A_{\mu}$ and $\widetilde{F}^{\mu\nu}=(1/2)\epsilon^{\mu\nu\rho\sigma}F_{\rho\sigma}$ is
its dual. Such a rotation is independent of the frequencies, different
from the Faraday rotation experienced by the electromagnetic waves when they pass through a magnetic field.
This phenomenon may be observable on large scales, for instances on
astrophysical or cosmological scales a tiny coupling might induce a significant rotation angle potentially be
observable by the current or near future experiments. The Chern-Simons coupling violates Lorentz and CPT symmetries
in the background, but the gauge symmetry is preserved
in cases when $p_{\mu}$ is constant or a derivative of a scalar field, i.e., $p_{\mu}\propto \partial_{\mu}\phi$.
In cosmology such a rotation will change the angular spectra of the
polarization of the cosmic microwave background radiation (CMB), especially rotate partially the $E$ mode polarization into
the $B$ mode one \cite{lue}. Since both $E$ and $B$ modes are changed, we can use the high quality CMB observational data
to detect the rotation angle even though no explicit $B$ mode signal has been detected. This opened a new window
to test the fundamental symmetries by the cosmological probe and since the first work \cite{cptv_boomrang} it has stimulated many
research interests in this field \cite{Liu:2006uh}. Currently the contraint on the
homogeneous rotation angle from the Wilkinson Microwave Anisotropy Probe (WMAP) 9-year data
is $\bar{\chi}=-0.36\pm 1.24\pm 1.5$ degree \cite{wmap9}.

If the external field $p_{\mu}$ is a constant vector, there will be inconsistencies between the electromagnetic and gravitational field
equations if the law of gravity is governed by Einstein's general relativity or its covariant extensions \cite{Li:2009rt}.
Alternatively the external field may be originated from derivatives of some scalar fields, e.g., $p_{\mu}=(c/M)\partial_{\mu}\phi$ with
$M$ a certain mass scale which indicates the cutoff energy scale of the effective theory and
$c$ the dimensionless coupling constant.
It is present in models where the scalar field $\phi$ is a pseudo Goldstone boson produced
by $U(1)$ symmetry breaking. Like Axion, this scalar field has the shift symmetry and only couples to
other matter derivatively. It is expected to interact with the electromagnetic field naturally via the Chern-Simons term.
The field $\phi$  may take the role of dark energy \cite{carroll} or
not \cite{pospelov}. It was also motivated by the leptogenesis models \cite{Li:2006ss} in which the baryon number asymmetry is produced
in thermal equilibrium \cite{quin_baryogenesis}, this violates one of the Sakharov's conditions \cite{Sakharov:1967dj}
obeyed in traditional baryogenesis models because
$CPT$ symmetry is violated by the evolution of $\phi$ along the time direction \cite{Cohen:1987vi}.
In this case all the equations are consistent, but the rotation angle which depends on the scalar field is not homogeneous
across the sky because the scalar field, as a dynamical component, has different values at different positions.
The fluctuations, treated as small effects, will bring new distortions to the CMB spectra \cite{Li:2008tma}
very similar to the effects of weak lensing on CMB \cite{Lewis:2006fu}.
Measuring the direction dependence of the rotation angle with the CMB data is important to investigate the dynamics of the new
scalar field $\phi$. Up to now the CMB constraint on the power spectrum of the rotation angle from WMAP 7-year data was put
in Ref.\cite{Gluscevic:2012me} based on the formalism developed in Refs. \cite{Kamionkowski:2008fp,Gluscevic:2009mm}. They obtained the result that for
the multipoles from $l=1$ to $l=512$, there is no evidence of non-zero power spectrum for the rotation angle within $3\sigma$.

In this paper we revisit this problem and compute the new constraints on the anisotropic rotation by analyzing
the recently released WMAP 9-year data combined with data from the ground based CMB experiments QUaD and BICEP. We will not adopt
any model for the scalar field.
We find that the dominant effect on the CMB spetra of the anisotropic rotation angle is through the suppression factors of its variance so the
current data put a strong constraint on the variance as well as isotropic part of the rotation angle. In comparison, current observational
data are not sensitive to the detailed shape of the power spectrum of the rotation angle.

This paper is organized as follows. In Sec. II, we
introduce the relevant formalism for computing the rotated power spectra of CMB. In Sec. III, we
present our result of the constraint on the homogeneous and anisotropic rotation angle from current
observational data. Sec. IV is the conclusion.

\section{Formalism of Rotated Power Spectra}

The full Lagrangian density of the Maxwell theory modified by the Chern-Simons term (without other sources) is
\be\label{lagrangian}
\mathcal{L}=-{1\over 4}F_{\mu\nu}F^{\mu\nu}+
p_{\mu}A_{\nu}\widetilde{F}^{\mu\nu}~.
\ee
This Lagrangian density is not gauge-invariant, but
the action integral $S=\int \mathcal{L} d^4x$ is gauge-independent
because in this paper we only consider $p_{\mu}=(c/M)\partial_{\mu}\phi$ and it is easy to see that after
integral by part the Chern-Simons term can be rewritten as $-(c\phi/2M)F_{\mu\nu}\widetilde{F}^{\mu\nu}$.

When applying the modified theory to CMB, the Chern-Simons term is expected to be significant only in the era after
recombination. Before that, CMB photons coupled tightly to the electrons and the Chern-Simons coupling is negligible compared
with the dominated Compton scattering.
After the last scattering off the electrons, the polarization direction of the photon rotated
by an angle \cite{Li:2008tma}
\be
\chi=-\int^{0}_{LSS}p_{\mu}dx^{\mu}(\lambda)=\frac{c}{M}\Delta\phi~,
\ee
where the integral along the light path (parameterized by $\lambda$) is from the light source to the point of
observation. For CMB, the light source is located at the last scattering surface and one can see that the rotation angle only relies on
the total difference of the field between the source and the observation point,
$\Delta\phi=\phi(\vec{x}_{LSS}, \eta_{LSS})-\phi_0$, where $\vec{x}_{LSS}$ represents the position on the last scattering surface and $\eta$
is the conformal time. The observation point is fixed but on the last scattering surface the scalar field
$\phi$ varies with the positions. In the spatially flat universe $\vec{x}_{LSS}=(\eta_0-\eta_{LSS})\hat{\bi{n}}$, the rotation angle relies
on the direction $\hat{\bi{n}}$ through the dependence of the field on the position.
Accordingly the Stokes
parameters of linear polarization got a rotation
\be\label{ro}
(\tilde{Q}\pm i\tilde{U})=\exp{(\pm
i2\chi)}(Q\pm iU)~,
\ee
where we have used tilde to denote the rotated parameter. More details can be found in Ref. \cite{Li:2008tma}.

On the full sky the temperature and polarization
fields can be decomposed in terms of appropriate spin-weighted harmonic
functions \cite{Zaldarriaga:1996xe}: \bea
T(\hat{\bi{n}})&=& \sum_{lm}a_{T,lm}Y_{lm}(\hat{\bi{n}})\nonumber \\
(Q\pm iU) (\hat{\bi{n}})&=& \sum_{lm} a_{\pm 2, lm} \;_{\pm 2}Y_{lm}(\hat{\bi{n}})~.
\eea
The expressions for the expansion coefficients are
\begin{eqnarray}
a_{T,lm}&=&\int d\Omega\; Y_{lm}^{*}(\hat{\bi{n}}) T(\hat{\bi{n}})
\nonumber  \\
a_{\pm 2,lm}&=&\int d\Omega \;_{\pm 2}Y_{lm}^{*}(\hat{\bi{n}}) (Q\pm iU)(\hat{\bi{n}})~.\label{alm}
\end{eqnarray}
Instead of $a_{2,lm}$ and $a_{-2,lm}$, it is more convenient to use their
linear combinations
\begin{eqnarray}
a_{E,lm}=-(a_{2,lm}+a_{-2,lm})/2 \nonumber \\
a_{B,lm}=i(a_{2,lm}-a_{-2,lm})/2.
\label{aeb}
\end{eqnarray}
The advantage of $E/B$ decomposition is that it is coordinate-independent and the $E$ and
$B$ modes represent polarization patterns of opposite parity.
The power spectra are defined as \be \langle a_{X',l^\prime
m^\prime}^{*} a_{X,lm}\rangle= C^{X'X}_{l} \delta_{l^\prime l}
\delta_{m^\prime m} \ee with the assumption of statistical isotropy.
In the equation above, $X'$ and $X$ denote the temperature $T$ and
the $E$ and $B$ modes of the polarization field, respectively. For Gaussian
theories, the statistical properties of the CMB
temperature/polarization maps are specified fully by these six
spectra. Without the rotation induced by the Chern-Simons coupling, $C^{TB}_l=C^{EB}_l=0$.

The Chern-Simons term has no effect on the temperature field. The rotated polarization field (\ref{ro})
can be decomposed in the same way except the coefficients and the spectra should be denoted by tildes.
We assume the field $\phi$ which induces the rotation is also a Gaussian random variable, as we have done for studies of cosmic scalar fields
in other cases. So we also have a power spectrum for the rotation angle.
As usual, we separate the rotation angle into its background and fluctuation,
\be
\chi(\hat{\bi{n}})=\bar{\chi}+\delta\chi(\hat{\bi{n}})~,
\ee
the background part $\bar{\chi}$ is homogeneous across the sky and the perturbation $\delta\chi(\hat{\bi{n}})$
as a scalar
with zero mean can be decomposed by the spherical harmonics,
\be
\delta\chi(\hat{\bi{n}})=\sum_{lm}b_{lm}Y_{lm}(\hat{\bi{n}})~.
\ee
With
assumed statistical isotropy of $b_{lm}$, the angular power spectrum of the rotation angle is defined as
\be\label{chi3} \langle
b_{l^\prime m^\prime}^{*} b_{lm}\rangle= C^{\chi}_{l}
\delta_{l^\prime l} \delta_{m^\prime m} ~,
\ee
which depends on the three dimensional power spectrum of the scalar field $\phi$ at the last scattering surface,
\be
C_l^{\chi}=\frac{4\pi c^2}{M^2}\int d\ln k {\cal P}_{\phi}(k,~\eta_{LSS}) j_l^2(k\Delta\eta)~,
\ee
where $\Delta\eta=\eta_0-\eta_{LSS}$ and $j_l$ is the spherical Bessel function.
With these equations we can calculate the rotated spectra analytically under the expansion of the factor
$\exp{(\pm 2i\chi)}$ into Taylor series by assuming the rotation angle is small.
Up to the second order (for spectra the fourth order), the rotated
spectra of CMB was shown explicitly in Eq. (69) in Ref. \cite{Li:2008tma}.

In this paper, instead of using Taylor expansion, we
will perform the non-perturbative calculations of the rotated power spectra via the computations of the rotated
correlation functions. This method has been used to calculate the lensing effect of CMB on all
scales \cite{Seljak:1995ve,Lewis:2006fu}, and it was shown that it is more accurate than the approach of Taylor expansion.
The correlation function only depends on the separation of the two points and is invariant under displacements.
In the spherical coordinate system with the origin was set to the position of the observer,
we need only evaluate the following three rotated correlation functions for the polarization
field by taking $\hat{\bi{n}}$ along the $z-$axis and $\hat{\bi{n'}}$
in the $x-z$ plane at angle $\beta$ to the $z-$axis
\bea
& &\tilde{\xi}_+\equiv \langle (\tilde{Q}+i\tilde{U})^{\ast}(\hat{\bi{n}})(\tilde{Q}+i\tilde{U})(\hat{\bi{n'}})\rangle\nonumber\\
& &\tilde{\xi}_-\equiv\langle (\tilde{Q}+i\tilde{U})(\hat{\bi{n}})(\tilde{Q}+i\tilde{U})(\hat{\bi{n'}})\rangle\nonumber\\
& &\tilde{\xi}_X\equiv\langle T(\hat{\bi{n}})(\tilde{Q}+i\tilde{U})(\hat{\bi{n'}})\rangle~.
\eea
They are related to the power spectra as
\bea
\tilde{\xi}_+(\beta)
&=&\sum_{lm,l'm'}\langle(\tilde{a}^{\ast}_{E,lm}-i\tilde{a}^{\ast}_{B,lm})(\tilde{a}_{E,l'm'}+i\tilde{a}_{B,l'm'})\rangle
_{2}Y_{lm}^{*}(\hat{\bi{n}})
_{2}Y_{l'm'}(\hat{\bi{n'}})\nonumber\\
&=&\sum_{lm}(\tilde{C}_l^{EE}+\tilde{C}_l^{BB})_{2}Y_{lm}^{*}(\hat{\bi{n}})
_{2}Y_{lm}(\hat{\bi{n'}})\nonumber\\
&=&\sum_l\frac{2l+1}{4\pi}(\tilde{C}_l^{EE}+\tilde{C}_l^{BB})d^l_{22}(\beta)~,\label{co1}
\eea
and
\bea
\tilde{\xi}_-(\beta)
&=&\sum_l\frac{2l+1}{4\pi}(\tilde{C}_l^{EE}-\tilde{C}_l^{BB}+2i\tilde{C}_l^{EB})d^l_{-22}(\beta)~,\label{co2}\\
\tilde{\xi}_X(\beta)
&=&-\sum_l\frac{2l+1}{4\pi}(\tilde{C}_l^{TE}+i\tilde{C}_l^{TB})d^l_{02}(\beta)~,\label{co3}
\eea
where $d^l_{mk}(\beta)$ is the Wigner small $d$ function
and $\cos\beta=\hat{\bi{n}}\cdot  \hat{\bi{n'}}$. We have included the rotated $TB$ and $EB$ correlations.
The reversals of the formulae (\ref{co1}), (\ref{co2}) and (\ref{co3}) give rise to
\bea\label{co}
& &\tilde{C}_l^{EE}+\tilde{C}_l^{BB}=2\pi \int^1_{-1}\tilde{\xi}_+(\beta) d^l_{22}(\beta) d\cos\beta\nonumber\\
& &\tilde{C}_l^{EE}-\tilde{C}_l^{BB}=2\pi \int^1_{-1}{\rm Re}[\tilde{\xi}_-(\beta)] d^l_{-22}(\beta) d\cos\beta\nonumber\\
& &\tilde{C}_l^{EB}=\pi \int^1_{-1}{\rm Im}[\tilde{\xi}_-(\beta)] d^l_{-22}(\beta) d\cos\beta\nonumber\\
& &\tilde{C}_l^{TE}=-2\pi \int^1_{-1}{\rm Re}[\tilde{\xi}_X(\beta)] d^l_{02}(\beta) d\cos\beta\nonumber\\
& &\tilde{C}_l^{TB}=-2\pi \int^1_{-1}{\rm Im}[\tilde{\xi}_X(\beta)] d^l_{02}(\beta) d\cos\beta~.
\eea

Furthermore, according to the Eq. (\ref{ro}), the rotated correlation function is related to the unrotated
correlation function as
\bea\label{co4}
\tilde{\xi}_+(\beta)&=&\langle \exp{[2i(\chi(\hat{\bi{n'}})-\chi(\hat{\bi{n}}))]}(Q+iU)^{\ast}(\hat{\bi{n}})(Q+iU)(\hat{\bi{n'}})\rangle
\nonumber\\
&=& \langle \exp{[2i(\chi(\hat{\bi{n'}})-\chi(\hat{\bi{n}}))]}\rangle \xi_+(\beta)\nonumber\\
&=&\exp{[-2\langle [\delta\chi(\hat{\bi{n'}})-\delta\chi(\hat{\bi{n}})]^2\rangle]}\xi_+(\beta)\nonumber\\
&=& \exp{[-4C^{\chi}(0)+4C^{\chi}(\beta)]}\xi_+(\beta)\nonumber\\
&=&\exp{[-4C^{\chi}(0)+4C^{\chi}(\beta)]}
\sum_{l}\frac{2l+1}{4\pi}(C_l^{EE}+C_l^{BB})d^l_{22}(\beta)~,
\eea
where we have neglected the small effect of the correlation between the rotation angle and the unrotated polarization field and in the
third step we used the formula that
\be
\langle e^{ix} \rangle=e^{-\langle x^2 \rangle/2}
\ee
for a Gaussian variable $x$ (here it is $\delta\chi$) with zero mean. The notation $C^{\chi}(\beta)$
represents the two point correlation function of the perturbed rotation angle, i.e.,
\be
C^{\chi}(\beta)=\langle\delta\chi(\hat{\bi{n}})\delta\chi(\hat{\bi{n'}})\rangle=\sum_l\frac{2l+1}{4\pi}C_l^{\chi}P_l(\cos\beta)~,
\ee
and $C^{\chi}(0)=\sum_l\frac{2l+1}{4\pi}C_l^{\chi}$ is its variance.
Substitute Eq.(\ref{co4}) into the first equation of (\ref{co}),
we have
\bea\label{spectra1}
\tilde{C}_l^{EE}+\tilde{C}_l^{BB}=e^{-4C^{\chi}(0)}\sum_{l'}\frac{2l'+1}{2}(C_{l'}^{EE}+C_{l'}^{BB})\int^1_{-1}d^{l'}_{22}(\beta)d^l_{22}(\beta)
e^{4C^{\chi}(\beta)}d\cos\beta~.
\eea
Straightforwardly one may obtain that
\bea\label{powerspectra}
& &\tilde{C}^{EE}_l-\tilde{C}^{BB}_l=\cos(4\bar{\chi})e^{-4C^{\chi}(0)}
\sum_{l'}\frac{2l'+1}{2}(C_{l'}^{EE}-C_{l'}^{BB})\int^1_{-1}d^{l'}_{-22}(\beta)d^l_{-22}(\beta)e^{-4C^{\chi}(\beta)}d\cos\beta\nonumber\\
& &\tilde{C}^{EB}_l=\sin(4\bar{\chi})e^{-4C^{\chi}(0)}
\sum_{l'}\frac{2l'+1}{4}(C_{l'}^{EE}-C_{l'}^{BB})\int^1_{-1}d^{l'}_{-22}(\beta)d^l_{-22}(\beta)e^{-4C^{\chi}(\beta)}d\cos\beta\nonumber\\
& &\tilde{C}^{TE}_l=\cos(2\bar{\chi})e^{-2C^{\chi}(0)}
\sum_{l'}\frac{2l'+1}{2}C_{l'}^{TE}\int^1_{-1}d^{l'}_{02}(\beta)d^l_{20}(\beta)d\cos\beta=C_l^{TE}\cos(2\bar{\chi})e^{-2C^{\chi}(0)}\nonumber\\
& &\tilde{C}^{TB}_l=\sin(2\bar{\chi})e^{-2C^{\chi}(0)}
\sum_{l'}\frac{2l'+1}{2}C_{l'}^{TE}\int^1_{-1}d^{l'}_{02}(\beta)d^l_{20}(\beta)d\cos\beta=C_l^{TE}\sin(2\bar{\chi})e^{-2C^{\chi}(0)}~.
\eea
Now we can use these formulae to compute the rotated power spectra and use the observational data together with the MCMC
package to search for or put the constraints on $\bar{\chi}$ and $C_l^{\chi}$.

Before that we note the spectra $\tilde{C}_l^{EE}$, $\tilde{C}_l^{BB}$ and $\tilde{C}_l^{EB}$ may be decomposed into the sums,
\be\label{decomposition}
\tilde{C}_l^{EE}=\tilde{C}_{l,0}^{EE}+\Delta \tilde{C}_l^{EE}~,~~\tilde{C}_l^{BB}=\tilde{C}_{l,0}^{BB}+\Delta \tilde{C}_l^{BB}~,
~~\tilde{C}_l^{EB}=\tilde{C}_{l,0}^{EB}+\Delta \tilde{C}_l^{EB}~,
\ee
where
\bea\label{decomposition1}
& &\tilde{C}_{l,0}^{EE}=[C_l^{EE}\cos^2(2\bar{\chi})+C_l^{BB}\sin^2(2\bar{\chi})]e^{-4C^{\chi}(0)}\nonumber\\
& &\tilde{C}_{l,0}^{BB}=[C_l^{EE}\sin^2(2\bar{\chi})+C_l^{BB}\cos^2(2\bar{\chi})]e^{-4C^{\chi}(0)}\nonumber\\
& &\tilde{C}_{l,0}^{EB}={1\over 2} \sin(4\bar{\chi})(C_l^{EE}-C_l^{BB})e^{-4C^{\chi}(0)}
\eea
together with $\tilde{C}_l^{TE}$ and $\tilde{C}_l^{TB}$ they depend on the power spectrum of the rotation angle
$C_l^{\chi}$ only through the variance $C^{\chi}(0)$.
Other parts rely on the correlation function of nonzero angular separation,
\bea\label{decomposition2}
& &\Delta\tilde{C}_l^{EE}+\Delta\tilde{C}_l^{BB}=2\pi\int^1_{-1}\Delta\tilde{\xi}_{+}(\beta)d_{22}^l(\beta)d\cos\beta\nonumber\\
& &\Delta\tilde{C}_l^{EE}-\Delta\tilde{C}_l^{BB}=2\pi\int^1_{-1}{\rm Re}[\Delta\tilde{\xi}_{-}(\beta)]d_{22}^l(\beta)d\cos\beta\nonumber\\
& &\Delta\tilde{C}_l^{EB}=\pi\int^1_{-1}{\rm Im}[\Delta\tilde{\xi}_{-}(\beta)]d_{22}^l(\beta)d\cos\beta~,
\eea
where
\bea
& &\Delta\tilde{\xi}_{+}(\beta)=e^{-4C^{\chi}(0)}(e^{4C^{\chi}(\beta)}-1)\sum_l\frac{2l+1}{4\pi}(C_l^{EE}+C_l^{BB})d^l_{22}(\beta)\nonumber\\
& &\Delta\tilde{\xi}_{-}(\beta)=e^{4(i\bar{\chi}-C^{\chi}(0))}(e^{-4C^{\chi}(\beta)}-1)\sum_l\frac{2l+1}{4\pi}(C_l^{EE}-C_l^{BB})d^l_{-22}(\beta)
~.
\eea

\section{Results}

In the above section, we have calculated the rotated CMB power spectra based on the full sky formalism. 
The results showed that besides the effects of homogeneous rotation $\bar{\chi}$, the anisotropic rotation angle brought new 
distortions to the CMB spectra, indicated by the variance $C^{\chi}(0)$ and two point correlation function $C^{\chi}(\beta)$ in the formulae (\ref{spectra1}) and (\ref{powerspectra}), 
both of which depend on the angular power spectrum $C_l^{\chi}$ and in principle can be constrained by the data from CMB polarization experiments. 
In order to constrain the homogeneous and anisotropic rotation angle in a model independent way, that means we will not assume any model for the 
scalar field $\phi$, 
we first modify the publicly available MCMC package CosmoMC \cite{Lewis:2002ah}
to include the distortions in Eqs. (\ref{spectra1}) and (\ref{powerspectra}) by treating $\bar{\chi}$ as a free parameter and 
$C_l^{\chi}$ as a positive otherwise undetermined function of $l$, and then
perform a global fit to the CMB data.
This method is different from that developed in \cite{Kamionkowski:2008fp,Yadav:2009eb,Gluscevic:2012me},
which adopted a quadratic estimator to constrain the anisotropic rotation angle.
The main difference is that in our method
we treat the rotation angle as a Gaussian random field, its statistics is isotropic, 
so that the rotated polarizations of CMB which were identified as the observables are also statistically isotropic and 
there will be no off-diagonal correlations between different multipoles $l$ and $l'$.   
However, the quadratic estimator
method considers the ensemble of CMB rotated by a fixed rotation angle $\chi(\hat{\bi n})$ which is not averaged and breaks the statistical isotropy of the 
rotated polarization field of CMB. Hence
the off-diagonal modes of the CMB correlations do not vanish and
depend explicitly on the rotation angle $\chi(\hat{\bi n})$,
since the averaging is done only on the CMB ensemble.
Especially, in the quadratic estimator method, some of the CMB correlations contain $O (\delta \chi)$
terms, based on which the quadratic estimator is constructed.
Different from this, in our method both the unrotated CMB field and the 
rotation field are considered as Gaussian random fields and have isotropic statistics. 
The averaging are done on both the CMB and
rotation field ensembles, only the diagonal terms are nonzero
and the results do not depend on $\delta\chi$ in the coordinate space, but rather on its auto-correlation functions
(or its power spectrum $C_l^{\chi}$).
This point can be understood in detail by comparing the CMB $EB$ correlation appearing
in \cite{Kamionkowski:2008fp} (equation (9)) and that
in our paper. The CMB $EB$ correlation in \cite{Kamionkowski:2008fp}
has off-diagonal modes, and its first term is $O (\bar{\chi})$,
while the second term is $O (\delta \chi)$. In our paper,
the CMB $EB$ correlation only has diagonal $\tilde{C}_l^{EB}$ modes
and only $O (\bar{\chi})$ and $O (C_l^{\chi})$
terms appear in $\tilde{C}_l^{EB}$ (the $O (\bar{\chi})$ term
is the same as that in \cite{Kamionkowski:2008fp} if $\bar{\chi}$
is replaced by $\frac{\alpha_{00}}{\sqrt{4\pi}}$). So in our analysis it is necessary to model the shape of $C_l^{\chi}$, in the quadratic estimator method it does not need to do so.

In our computations, besides the WMAP 9yr temperature and polarization data
\cite{Bennett:2012fp} (including the low-$l$ TB, low-$l$ EB and high-$l$ TB data),
we also adopt the QUaD data \cite{Brown:2009uy} and
the BICEP data \cite{Chiang:2009xsa}. As for the cosmological
model, we use a flat $\Lambda \text{CDM}$ Universe with a cosmological constant.
The free cosmological parameters are $(\omega_b, \omega_c, \Theta_s, \tau, n_s, r, A_s)$,
where $\omega_b \equiv \Omega_bh^2$ and $\omega_c \equiv \Omega_{c}h^2 $
are the baryon density and cold dark matter density respectively, $\Theta_s$ is the ratio of
the sound horizon to the angular diameter distance at decoupling,
$\tau$ is the optical depth to reionization, $n_s$ is scalar spectral
index, $r$ is the tensor to scalar ratio of the primordial spectrum
and $A_s$ defines the amplitude of the primordial scalar spectrum.

\begin{figure}
\begin{center}
\includegraphics[width=1.0\textwidth]{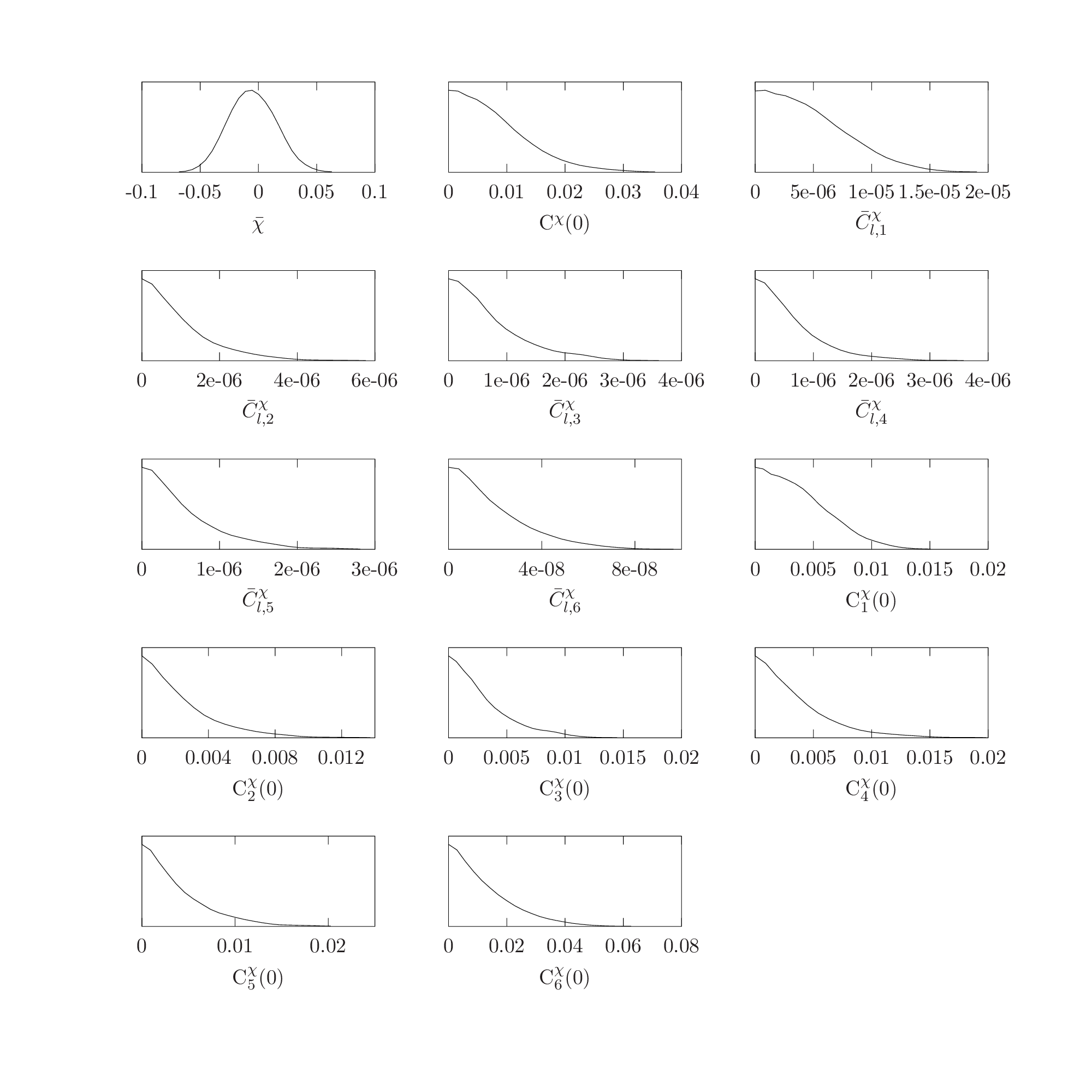}
\end{center}
\caption{One-dimensional constraints on the isotropic rotation angle $\bar{\chi}$,
the variance of the anisotropies of the rotation angle $C^{\chi}(0)$,
binned $C_l^{\chi}$ parameters $\bar{C}^{\chi}_{l,i}$ and derived parameters
$C_i^{\chi}(0)$. In obtaining this result, we let $C^{\chi}(0)$
runs freely with the binned $C_l^{\chi}$ parameters.}
\label{p_chib2}
\end{figure}

Compared with the choice for the cosmological parameters, it is difficult to
choose the parameters for the anisotropic rotation angle.
From Eqs. (\ref{spectra1}, \ref{powerspectra}, \ref{decomposition}, \ref{decomposition1}, \ref{decomposition2}) we note that
the distortions to the CMB power spectra brought by the anisotropies of
the rotation angle can be divided into two parts. The first part appears
in $\tilde{C}_{l,0}^{EE},\tilde{C}_{l,0}^{BB},\tilde{C}_{l,0}^{EB}$, $\tilde{C}_l^{TE}$ and $\tilde{C}_l^{TB}$,
in which the anisotropies of the rotation angle suppress the CMB angular power
spectra through the variance $C^{\chi}(0)$ via the factors
$\exp(-4C^{\chi}(0))$ and $\exp(-2C^{\chi}(0))$. The second one is
$\Delta\tilde{C}_l^{EE}$, $\Delta\tilde{C}_l^{BB}$
and $\Delta\tilde{C}_l^{EB}$, which depends on the detailed shape of
$C_l^{\chi}$. Generally speaking, the first one dominates the CMB power spectra distortions.
Therefore, the data will give a strong constraint
on $C^{\chi}(0)$ but a weak constraint on the
shape of $C_l^{\chi}$. Furthermore, the constraint on $C^{\chi}(0) = \sum_l\frac{2l+1}{4\pi}C_l^{\chi}$
coming directly from the $\tilde{C}_{l,0}^{EE},\tilde{C}_{l,0}^{BB},
\tilde{C}_{l,0}^{EB}$, $\tilde{C}_l^{TE}$ and $\tilde{C}_l^{TB}$
terms is larger than that coming directly from
the $\Delta\tilde{C}_l^{EE}$, $\Delta\tilde{C}_l^{BB}$
and $\Delta\tilde{C}_l^{EB}$ terms. Due to this, it is difficult to obtain
a precise constraint on the shape of $C_l^{\chi}$. We can only get a loose constraint on
it. The method we adopted is to let $C^{\chi}(0)$
runs freely with $C_l^{\chi}$. More specifically, we adopt six binned $C_l^{\chi}$ parameters
$\bar{C}^{\chi}_{l,1}, \bar{C}^{\chi}_{l,2}, \bar{C}^{\chi}_{l,3}, \bar{C}^{\chi}_{l,4}, \bar{C}^{\chi}_{l,5},\bar{C}^{\chi}_{l,6}$,
which are the average values of $C_l^{\chi}$ in
the multipole regions $[2, 100], [101,200],[201,300],[301,400],[401,500],[501,l_{max}]$
respectively ($l_{max} = 2550$ is the largest multipole of $C_l^{\chi}$
we used), to model the shape of $C_l^{\chi}$. So in this case, the general parameter space is
\bea
P = \{ \omega_b, \omega_c, \Theta_s, \tau, n_s, r, A_s, \bar{\chi},
C^{\chi}(0), \bar{C}^{\chi}_{l,1}, \bar{C}^{\chi}_{l,2}, \bar{C}^{\chi}_{l,3}, \bar{C}^{\chi}_{l,4}, \bar{C}^{\chi}_{l,5},
\bar{C}^{\chi}_{l,6} \}.
\eea
Besides these quantities we also calculate six derived parameters
$C_i^{\chi}(0)\equiv \sum_{l=l_{i,min}}^{l=l_{i,max}}\frac{2l+1}{4\pi}\bar{C}^{\chi}_{l,i}$
where $i=1-6$, which can be used to compare $C^{\chi}(0)$ with
$\sum_{i=1}^6C^{\chi}_i(0)$. When running the program, we set a physical prior on
$C^{\chi}(0)$ and $\bar{C}^{\chi}_{l,i}$, that is $C^{\chi}(0)>0$
and $\bar{C}^{\chi}_{l,i}>0$, but we do not require $C^{\chi}(0)\geq \sum_{i=1}^6C^{\chi}_i(0)$.
The constraints on these parameters are illuminated in Figure \ref{p_chib2},
from which no positive signal of nonzero
CMB polarization rotation or possible anisotropies are found,
but a strong constraint is obtained. More specifically,
at 2$\sigma$ level of confidence, the constraints are
$-0.043 <  \bar{\chi}  < 0.034$ (throughout this section, radian
is adopted as the unit of the angle), $C^{\chi}(0)  <  0.020$,
$\bar{C}_{l,1}^{\chi}  < 1.1 \times 10^{-5}$, $\bar{C}_{l,2}^{\chi}  < 2.8 \times 10^{-6}$,
$\bar{C}_{l,3}^{\chi}  < 2.0 \times 10^{-6}$, $\bar{C}_{l,4}^{\chi}  < 1.8 \times 10^{-6}$,
$\bar{C}_{l,5}^{\chi}  < 1.5 \times 10^{-6}$, $\bar{C}_{l,6}^{\chi}  < 5.2 \times 10^{-8}$,
$C^{\chi}_1(0)  <  0.0090$, $C^{\chi}_2(0)  <  0.0068$, $C^{\chi}_3(0)  <  0.0081$, $C^{\chi}_4(0)  <  0.0099$,
$C^{\chi}_5(0)  <  0.011$ and $C^{\chi}_6(0)  <  0.034$ respectively.
Compared with the results presented in \cite{Gluscevic:2012me}, it seems that the constraint
obtained here is tighter. However, it is difficult to directly compare them
with each other, since the two methods differ in several aspects.
First, as we mentioned at the beginning of this section, we treated the anisotropic rotation angle as a 
Gaussian random field with isotropic statistics
and use the rotated power spectra to constrain its anisotropies, while in \cite{Gluscevic:2012me} the authors adopted 
the quadratic maps to estimate the fixed anisotropic rotation angle. Second,
we set a positive prior on the $C^{\chi}(0)$ and $C_l^{\chi}$. Third, the bin size
we adopted is $\Delta l \sim 100$, which is different from $\Delta l \sim 50$
used in \cite{Gluscevic:2012me}. 

\begin{figure}
\begin{center}
\includegraphics[width=0.7\textwidth]{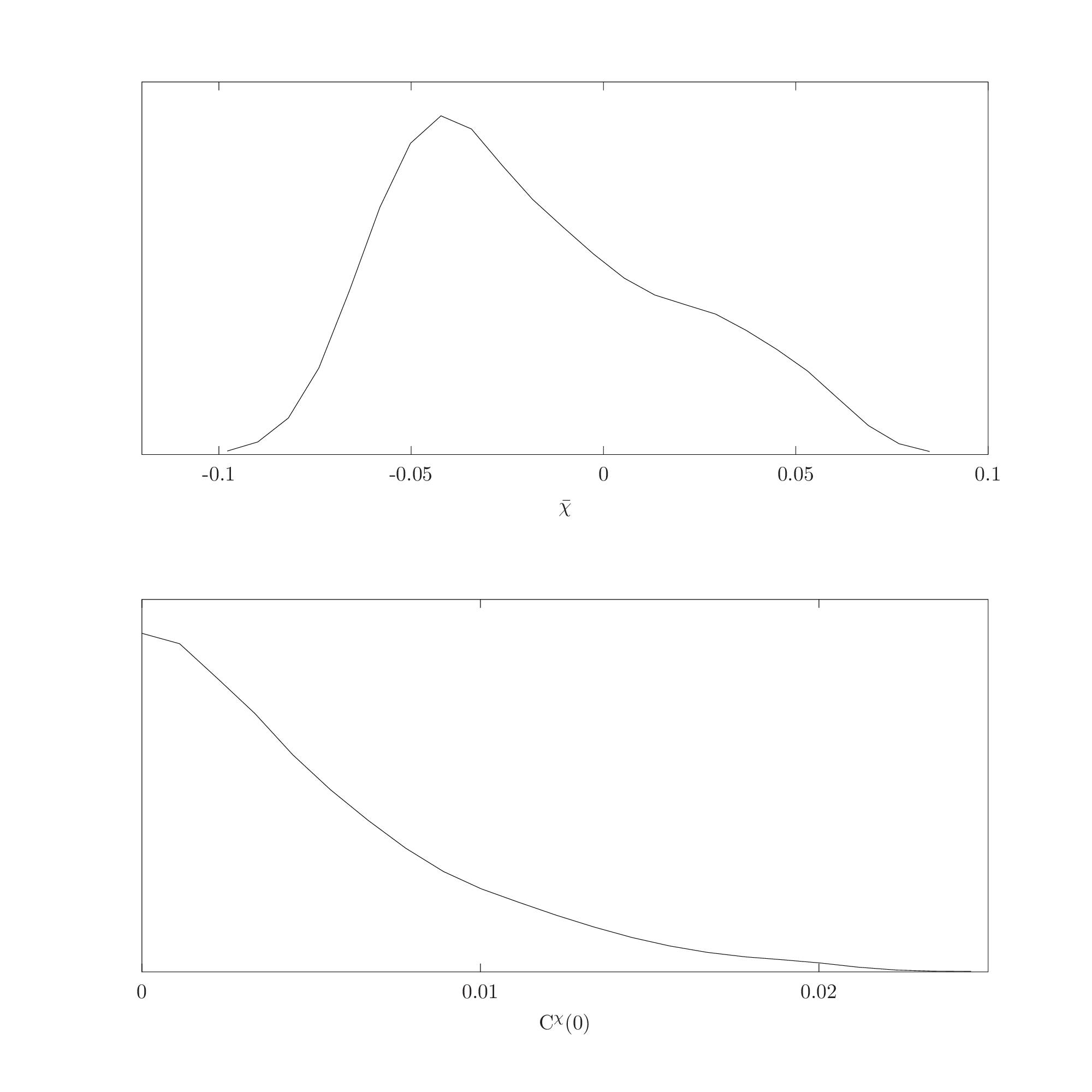}
\end{center}
\caption{One-dimensional constraints on the isotropic rotation angle $\bar{\chi}$ and
the variance of the anisotropies of the rotation angle $C^{\chi}(0)$. In obtaining this result,
we ignored the effects of $\Delta\tilde{C}_l^{EE}$, $\Delta\tilde{C}_l^{BB}$
and $\Delta\tilde{C}_l^{EB}$, which depend on the detailed shape of $C_l^{\chi}$.}
\label{p_chic}
\end{figure}

At the same time, one can see that
the constraint on $C^{\chi}(0)$ is much stronger than $\sum_{i=1}^{6} C_i^{\chi}(0)$, which
indicates that the distortion of the CMB power spectra induced by
the anisotropies of the rotation angle is indeed dominated by
the variance $C^{\chi}(0)$ which appears as suppression factors in $\tilde{C}_{l,0}^{EE},\tilde{C}_{l,0}^{BB},\tilde{C}_{l,0}^{EB}$,
$\tilde{C}_l^{TE}$ and $\tilde{C}_l^{TB}$.
Due to this, we consider another simpler way, in which
the effects of $\Delta\tilde{C}_l^{EE}$, $\Delta\tilde{C}_l^{BB}$
and $\Delta\tilde{C}_l^{EB}$ are ignored and the parameter space becomes
\bea
P=\{ \omega_b, \omega_c, \Theta_s, \tau, n_s, r, A_s, \bar{\chi},C^{\chi}(0) \}.
\eea
The results are displayed in Figure \ref{p_chic} and the 2$\sigma$ constraints
are $-0.071 <  \bar{\chi}  < 0.057$, $C^{\chi}(0)  <  0.014$. The constraints are
similar to that obtained with the first way, but the constraint on $\bar{\chi}$
becomes weaker and the constraint on $C^{\chi}(0)$ becomes stronger.

Based on the above analysis, we obtain the following conclusions about the results:
(1) Currently the data put strong constraints on isotropic rotation angle $\bar{\chi}$ as well as
the variance of the anisotropic rotation angle $C^{\chi}(0)$. We can conclude conservatively that $C^{\chi}(0)  <  0.020$;
(2) Although a precise constraint on the shape of $C_l^{\chi}$ is
not obtained, one can still give some weak constraints on it.

\section{Conclusions}

Detecting the rotation of the CMB polarization induced by the Chern-Simons coupling is now considered as an important way to test
Lorentz and CPT symmetries in the community. Usually people only focus on the isotropic rotation.
Theoretically if the rotation is induced by a fixed vector, when including the gravity, the equations of motions are not consistent.
Alternatively if it is induced by
the scalar field coupling, the rotation angle must be direction dependent and we should include the anisotropy of the rotation angle
in the more consistent analysis. Furthermore, measuring the anisotropy of the rotation angle is equivalent to determine the distribution of
the scalar field on the last scattering surface and this is very important for us to probe the detailed dynamics of the scalar field.
In this paper, in a model independent way, we studied the constraints on the anisotropic rotation angle by analyzing the recently released
WMAP 9-year data and the
data from QUaD and BICEP experiments. We found that even though current experiments have little to say about the shape of the power
spectrum of the rotation angle, but they put strong constraint on its variance and its mean value (the homogeneous part of the rotation
angle). Our result showed no significant evidence for a nonzero rotation angle.

\section{Acknowledgement}

The author ML is supported in part by National Science Foundation of China under Grants No. 11075074 and No. 11065004, by the Specialized
Research Fund for the Doctoral Program of Higher Education (SRFDP) under Grant No. 20090091120054 and by SRF for ROCS, SEM.
BY is supported in part by National Science Foundation of China under Grants No. 11103081 and 10973039.

{}

\end{document}